\documentclass[11pt]{article}

\usepackage{amsfonts,amssymb,chet,dsfont,graphicx,mathrsfs,pgfplots,tikz}

\newcommand{\1}{\mathds{1}}
\newcommand{\Op}[2]{\mathcal{O}_{#1}(\eta_{#2})}
\newcommand{\ee}[3]{(\eta_{#1}\cdot\eta_{#2})^{#3}}
\newcommand{\D}{\mathcal{D}}
\newcommand{\cOPE}[4]{{}_{#1}c_{#2#3}^{\phantom{#2#3}#4}}
\newcommand{\DOPE}[4]{{}_{#1}\mathcal{D}_{#2#3}^{\phantom{#2#3}#4}}
\newcommand{\Vev}[1]{\left\langle{#1}\right\rangle}

\title{Conformal Bootstrap in Embedding Space}

\author{Jean-Fran\c{c}ois Fortin$^{\ast,}$\email{jean-francois.fortin@phy.ulaval.ca} and Witold Skiba$^{\dagger,}$\email{witold.skiba@yale.edu}}

\affiliation{
$^\ast$D\'epartement de Physique, de G\'enie Physique et d'Optique\\Universit\'e Laval, Qu\'ebec, QC G1V 0A6, Canada\\
$^\dagger$Department of Physics, Yale University, New Haven, CT 06520, USA
}%Choices for affiliations $^{\ast,\dagger,\$,\S,\ddag,}$

\abstract{It is shown how to obtain conformal blocks from embedding space with the help of the operator product expansion.  The minimal conformal block originates from scalar exchange in a four-point correlation function of four scalars.  All remaining conformal blocks are simple derivatives of the minimal conformal block.  With the help of the orthogonality properties of the conformal blocks, the analytic conformal bootstrap can be implemented directly in embedding space, leading to a Jacobi-like definition of conformal field theories.}

\date{February 2016} %Uncomment this line for month to be fixed

\begin{document}

\maketitle

%\toc

%%%%%%%%%%%%%%%%%%%%%%%%%%%%%%%%%%%%%%%%%%%%%%%%%%
%%%%%%%%%%%%%%%%%%%%%%%%%%%%%%%%%%%%%%%%%%%%%%%%%%

\section{Introduction}\label{SecIntro}

Conformal field theories (CFTs) are the end-points of renormalization group flows.  As such, they could ultimately allow a complete classification of CFTs as relevant deformations of a subset of CFTs.  They also describe the statistical behavior of second-order phase transitions, where the correlation length diverges.  The importance of a deep understanding of CFTs cannot be overstated.

The additional symmetries of CFTs, due to the dilatation and special conformal generators, strongly constrain such theories.  For example, two- and three-point correlation functions are completely determined up to constant coefficients.  The use of crossing symmetry \cite{Ferrara:1973yt,Polyakov:1974gs} on four-point correlation functions leads to constraints on these coefficients. To implement this program it is necessary to know expressions for the conformal blocks, which are completely settled by conformal invariance.

A more fundamental quantity is the operator product expansion (OPE).  Indeed, conformal invariance implies that there exists an OPE which relates the product of two fields at different points to a sum over all fields at an arbitrary point \cite{Mack:1976pa}.  The OPE has profound consequences since it allows one to compute all $n$-point correlation functions in terms of OPE coefficients and two-point functions.  From the OPE, the conformal blocks represent the exchange of a particular field between the four initial fields, usually in the $s$-, $t$- or $u$-channel.

CFTs in $d$-dimensions  naturally live in a $(d+2)$-dimensional space called the embedding space \cite{Dirac:1936fq}.  The embedding space has been used to obtain the two- and three-point correlation functions (see \textit{e.g.} \cite{Weinberg:2010fx,Weinberg:2012mz}), but the OPE in embedding space has not been exploited to its full capacity (for pioneering work see \cite{Ferrara:1971zy,Ferrara:1971vh,Ferrara:1972cq,Ferrara:1973eg,Ferrara:1973yt}).  In this paper we describe the OPE in embedding space and use it to obtain the scalar conformal block for four-point correlation functions of four scalar fields.  The main motivation is carrying out the conformal bootstrap \cite{Ferrara:1973yt,Polyakov:1974gs} in embedding space (which is more natural as the conformal algebra acts linearly in embedding space) which is possible due to the fact that all conformal blocks can be obtained as appropriate derivatives of a unique entity: the scalar conformal block.

Combining the OPE in embedding space with the recent observation \cite{Isachenkov:2016gim} that conformal blocks are related to virtual Koornwinder polynomials \cite{ref1}, which are the $q$-deformed hyperbolic extensions of the Koornwinder polynomials \cite{Koornwinder197448,Koornwinder197459,Koornwinder1974357,Koornwinder1974370,doi:10.1137/0509028}, it suggests that a complete analytic conformal bootstrap can be implemented, leading to a Jacobi-like definition of CFTs solely in terms of the conformal data.\footnote{For work that revived the numerical conformal bootstrap, the reader is referred to \cite{Rattazzi:2008pe} and subsequent work.}  Indeed, the associated Fourier transform-like decomposition of the crossing symmetry equations should simply be the hyperbolic version of the Fourier series-like decomposition in terms of the orthogonal Koornwinder polynomials.  The output should therefore be an infinite sum of products of two OPE coefficients which ultimately vanishes, reminiscent of a non-compact version of the Jacobi identity.

This paper is organized as follows:  In Section \ref{SecOPE} the OPE in embedding space is described and the appropriate differential operator for scalar fields is obtained.  Section \ref{SecCF} discusses $n$-point correlation functions from the point of view of the OPE in embedding space.  We re-derive well-known results for scalar fields for $(n\leq4)$-point correlation functions.  In particular, we compute the scalar conformal block from the OPE in embedding space and discuss the conformal bootstrap in embedding space.  Finally, in Section \ref{SecConc} we consider the analytic conformal bootstrap from the point of view of the virtual Koornwinder polynomials, before discussing future work and concluding.  Throughout this paper, we use the notation of \cite{Ferrara:1971zy,Ferrara:1971vh,Ferrara:1972cq,Ferrara:1973eg,Ferrara:1973yt} for coordinates in embedding space, \textit{i.e.} $\eta^A$ is an embedding space coordinate.

%%%%%%%%%%%%%%%%%%%%%%%%%%%%%%%%%%%%%%%%%%%%%%%%%%
%%%%%%%%%%%%%%%%%%%%%%%%%%%%%%%%%%%%%%%%%%%%%%%%%%

\section{Operator Product Expansion}\label{SecOPE}

The OPE is the most fundamental element defining a CFT.  In embedding space, the OPE between two fields $\Op{i}{1}$ and $\Op{j}{2}$ can be written as
\eqn{\Op{i}{1}\Op{j}{2}=\sum_k\sum_{a=1}^{N_{ijk}}\cOPE{a}{i}{j}{k}\DOPE{a}{i}{j}{k}(\eta_1,\eta_2)\Op{k}{2}.}[EqOPE]
It is an infinite sum over fields $\Op{k}{2}$ and for each of these fields, there is a finite sum over the $N_{ijk}$ OPE coefficients $\cOPE{a}{i}{j}{k}$ with the appropriate differential operators $\DOPE{a}{i}{j}{k}(\eta_1,\eta_2)$.

The OPE in embedding space is constrained by consistency conditions on the embedding space light-cone $\eta^2=0$.  Moreover, the differential operators are prescribed by conformal invariance and by the irreducible Lorentz group representations of the fields.  They are simple derivatives of the embedding space positions which are well-defined on the embedding space light-cone.  Most importantly, the differential operators for non-scalar fields are straightforward generalizations of the differential operators when all fields in the OPE are scalars.  All the differential operators are completely determined and are ultimately built from a unique fundamental differential operator in embedding space.

Focusing on scalar fields $\phi_i$, the homogeneity property in embedding space (see \textit{e.g.} \cite{Ferrara:1971zy,Ferrara:1971vh,Ferrara:1972cq,Ferrara:1973eg,Ferrara:1973yt}) is
\eqn{\eta\cdot\partial\,\phi(\eta)=-\Delta_i\phi_i(\eta),}[EqHom]
where $\Delta_i$ is the conformal dimension of the scalar field $\phi_i$, fixes completely the sole differential operator (there is only one OPE coefficient when all fields are scalars) to be
\eqna{
\DOPE{}{i}{j}{k}(\eta_1,\eta_2)&=\frac{1}{\ee{1}{2}{\frac{1}{2}(\Delta_i+\Delta_j-\Delta_k)}}\D(\eta_1,\eta_2)^{-\frac{1}{2}(\Delta_i-\Delta_j+\Delta_k)},\\
\D(\eta_1,\eta_2)&=\ee{1}{2}{}\partial_2^2-\eta_1\cdot\partial_2(d-2+2\eta_2\cdot\partial_2).
}[EqDiffS]
The differential operator $\D(\eta_1,\eta_2)$ was introduced in \cite{Ferrara:1971zy,Ferrara:1971vh,Ferrara:1972cq,Ferrara:1973eg,Ferrara:1973yt}.  Indeed, for three scalar fields, there is only one OPE coefficient and the scalar differential operator \eqref{EqDiffS} is the unique (up to normalization) differential operator which acts coherently on the embedding space light-cone.  From \eqref{EqDiffS} it is straightforward to check that the OPE \eqref{EqOPE} satisfies the homogeneity property in embedding space \eqref{EqHom} for both coordinates $\eta_1$ and $\eta_2$.  How to obtain the unique fundamental differential operator which is the basis of all the differential operators $\DOPE{a}{i}{j}{k}(\eta_1,\eta_2)$ will be described elsewhere.

%%%%%%%%%%%%%%%%%%%%%%%%%%%%%%%%%%%%%%%%%%%%%%%%%%
%%%%%%%%%%%%%%%%%%%%%%%%%%%%%%%%%%%%%%%%%%%%%%%%%%

\section{Correlation Functions}\label{SecCF}

With the knowledge of the OPE in embedding space, it is straightforward to investigate the correlation functions.  Focusing on scalar fields, it will be shown how the technique put forward here naturally leads back to known results about $n$-point correlation functions. 

%%%%%%%%%%%%%%%%%%%%%%%%%%%%%%%%%%%%%%%%%%%%%%%%%%

\subsection{One- and Two-point Correlation Functions}\label{SSec12ptCF}

One-point correlation functions are trivial.  The only operator with non-vanishing one-point correlation function is the identity operator $\1$, which has a vanishing conformal dimension $\Delta_\1=0$ and for which $\Vev{\1}=1$.  Therefore, from the OPE \eqref{EqOPE}, it is clear that two-point correlation functions simply single out the identity operator $\1$ on the RHS of \eqref{EqOPE}.

For scalar fields, the associated differential operator in \eqref{EqDiffS} is thus
\eqn{\DOPE{}{i}{j}{\1}(\eta_1,\eta_2)=\frac{\delta_{\Delta_i\Delta_j}}{\ee{1}{2}{\frac{1}{2}(\Delta_i+\Delta_j)}}.}[EqDiffSTwoPTS]
Since the identity operator vanishes under the action of the differential operator \eqref{EqDiffS} unless $\Delta_i-\Delta_j+\Delta_\1=0$, the only non-trivial differential operator occurs when the scalar fields on the LHS have the same conformal dimensions,  $\Delta_i=\Delta_j$, which in turn implies \eqref{EqDiffSTwoPTS}.

From \eqref{EqOPE} and \eqref{EqDiffSTwoPTS}, the scalar two-point correlation function in embedding space is given by
\eqn{\Vev{\phi_i(\eta_1)\phi_j(\eta_2)}=\cOPE{}{i}{j}{\1}\frac{\delta_{\Delta_i\Delta_j}}{\ee{1}{2}{\Delta_i}}.}[EqTwoPTS]
Projecting onto position space, following \textit{e.g.} \cite{Weinberg:2010fx,Weinberg:2012mz}, gives the right scalar two-point function.  From Bose symmetry of scalar fields, the OPE coefficient $\cOPE{}{i}{j}{\1}$ is thus symmetric. Assuming the theory is unitary $\cOPE{}{i}{j}{\1}$ is also positive-definite.  Therefore, $\cOPE{}{i}{j}{\1}$ can play the role of a metric in field space and it is usually convenient to diagonalize it.

More generally, the action of conformal Casimir operators on two-point correlation functions of arbitrary fields shows that the only possible OPE with the identity operator on the RHS are the ones where both fields on the LHS of \eqref{EqOPE} have the same Casimir eigenvalues, up to a possible minus sign related to the two inequivalent spinor irreducible representations in even dimensions (depending if they are self-conjugate or not).  Group-theoretical arguments then imply that there is only one OPE coefficient for operators of any spin
\eqn{\Vev{\Op{i}{1}\Op{j}{2}}=\cOPE{}{i}{j}{\1}\DOPE{}{i}{j}{\1}(\eta_1,\eta_2),}[EqTwoPT]
generalizing the metric in field space to all fields.  In terms of conformal data, the two-point correlation functions are non-vanishing only if the conformal dimensions are the same as well as all the Dynkin indices of the irreducible Lorentz representations of the two fields, up to the possible interchange of the two spinor Dynkin indices in even dimensions (which is related to the aforementioned minus sign in the language of Casimir eigenvalues).

%%%%%%%%%%%%%%%%%%%%%%%%%%%%%%%%%%%%%%%%%%%%%%%%%%

\subsection{Three-point Correlation Functions}\label{SSec3ptCF}

With the help of the two-point correlation functions \eqref{EqTwoPT}, three-point correlation functions are directly obtained from \eqref{EqOPE}
\eqna{
\Vev{\Op{i}{1}\Op{j}{2}\Op{k}{3}}&=\sum_{k'}\sum_{a=1}^{N_{ijk'}}\cOPE{a}{i}{j}{k'}\DOPE{a}{i}{j}{k'}(\eta_1,\eta_2)\Vev{\Op{k'}{2}\Op{k}{3}}\\
&=\sum_{k'}\sum_{a=1}^{N_{ijk'}}\cOPE{a}{i}{j}{k'}\DOPE{a}{i}{j}{k'}(\eta_1,\eta_2)\cOPE{}{k'}{k}{\1}\DOPE{}{k'}{k}{\1}(\eta_2,\eta_3)\\
&=\sum_{a=1}^{N_{ijk}}{}_ac_{ijk}\,{}_a\mathcal{D}_{ijk}(\eta_1,\eta_2,\eta_3),
}[EqThreePT]
where in the first line the OPE was used between $\Op{i}{1}$ and $\Op{j}{2}$, while in the last line the result is written in a more symmetric way to match the symmetry of the LHS.  Note also that using the OPE on any two fields in the three-point correlation function must give the same result, which implies symmetry relations for the OPE coefficients and the differential operators once they are properly contracted with the metric and its corresponding differential operator.  This can be seen explicitly by focusing on three scalar fields.

It is straightforward to first compute the three-point correlation function of three scalar fields, which is given by
\eqna{
\Vev{\phi_i(\eta_1)\phi_j(\eta_2)\phi_k(\eta_3)}&=\sum_{k'}\cOPE{}{i}{j}{k'}\DOPE{}{i}{j}{k'}(\eta_1,\eta_2)\cOPE{}{k'}{k}{\1}\frac{\delta_{\Delta_k\Delta_{k'}}}{\ee{2}{3}{\Delta_k}}\\
&=\frac{c_{ijk}}{\ee{1}{2}{\frac{1}{2}(\Delta_i+\Delta_j-\Delta_k)}\ee{1}{3}{\frac{1}{2}(\Delta_i-\Delta_j+\Delta_k)}\ee{2}{3}{\frac{1}{2}(-\Delta_i+\Delta_j+\Delta_k)}},
}[EqThreePTS]
using \eqref{EqThreePT}, \eqref{EqTwoPTS} and \eqref{EqDiffS}.  This result follows from the identity
\eqn{\D(\eta_1,\eta_2)^q\ee{2}{3}{s}=(-2)^q(-s)_q(-s+1-d/2)_q\ee{1}{3}{q}\ee{2}{3}{s-q},}[EqDiff12e23]
defined on the embedding space light-cone where $(\alpha)_m=\Gamma(\alpha+m)/\Gamma(\alpha)$ is the Pochhammer symbol.  The identity \eqref{EqDiff12e23} is shown by direct computation for a positive integer $q$ and fractional calculus is used to extend the result to any real $q$.  Projecting \eqref{EqThreePTS} onto position space (see \textit{e.g.} \cite{Weinberg:2010fx,Weinberg:2012mz}) shows once again that this technique reproduces known results for scalar three-point correlation functions.

Bose symmetry of the three scalar fields in the three-point correlation function \eqref{EqThreePTS} implies that $c_{ijk}$ is fully symmetric under the interchange of its indices.  In terms of the OPE coefficients however, one has
\eqn{c_{ijk}=(-2)^{-\frac{1}{2}(\Delta_i-\Delta_j+\Delta_k)}(\Delta_k)_{-\frac{1}{2}(\Delta_i-\Delta_j+\Delta_k)}(\Delta_k+1-d/2)_{-\frac{1}{2}(\Delta_i-\Delta_j+\Delta_k)}\sum_{k'}\cOPE{}{i}{j}{k'}\cOPE{}{k'}{k}{\1},}
which demonstrates that the symmetry properties of $\cOPE{}{i}{j}{k}$ are not as simple (this extends to arbitrary OPE coefficients).

%%%%%%%%%%%%%%%%%%%%%%%%%%%%%%%%%%%%%%%%%%%%%%%%%%

\subsection{Four-point Correlation Functions}\label{SSec4ptCF}

Up to now the OPE in embedding space was used to obtain one-, two- and three-point correlation functions.  The results conform with the literature although they are somewhat trivial.  The real challenge is to reproduce the conformal blocks using the OPE in embedding space.

For arbitrary fields, the four-point correlation functions in the $s$-channel are
\eqna{
\Vev{\Op{i}{1}\Op{j}{2}\Op{k}{3}\Op{\ell}{4}}&=\sum_{m,n}\sum_{a=1}^{N_{ijm}}\sum_{b=1}^{N_{k\ell n}}\cOPE{a}{i}{j}{m}\cOPE{b}{k}{\ell}{n}\DOPE{a}{i}{j}{m}(\eta_1,\eta_2)\DOPE{b}{k}{\ell}{n}(\eta_3,\eta_4)\Vev{\Op{m}{2}\Op{n}{4}}\\
&=\sum_{m,n}\sum_{a=1}^{N_{ijm}}\sum_{b=1}^{N_{k\ell n}}\cOPE{a}{i}{j}{m}\cOPE{}{m}{n}{\1}\cOPE{b}{k}{\ell}{n}\DOPE{a}{i}{j}{m}(\eta_1,\eta_2)\DOPE{b}{k}{\ell}{n}(\eta_3,\eta_4)\DOPE{}{m}{n}{\1}(\eta_2,\eta_4).
}[EqFourPT]
The conformal blocks are simply obtained by isolating the appropriate exchanged field in the $s$-channel.

Concentrating on the scalar exchange in a four-point correlation function of four scalar fields, from \eqref{EqFourPT} the quantity of interest for the corresponding conformal block is
\eqna{
\tilde{G}&=\DOPE{}{i}{j}{m}(\eta_1,\eta_2)\DOPE{}{k}{\ell}{m}(\eta_3,\eta_4)\frac{1}{\ee{2}{4}{\Delta_m}}\\
&=\frac{1}{\ee{1}{2}{\frac{1}{2}(\Delta_i+\Delta_j-\Delta_m)}\ee{3}{4}{\frac{1}{2}(\Delta_k+\Delta_\ell-\Delta_m)}}\\
&\phantom{=}\hspace{1cm}\times\D(\eta_1,\eta_2)^{-\frac{1}{2}(\Delta_i-\Delta_j+\Delta_m)}\D(\eta_3,\eta_4)^{-\frac{1}{2}(\Delta_k-\Delta_\ell+\Delta_m)}\frac{1}{\ee{2}{4}{\Delta_m}}
.}[EqtG]
For scalar fields the remaining terms in \eqref{EqFourPT} only change the normalization of $\tilde{G}$.  Introducing the conformal ratios $u$ and $v$ which are defined as
\eqn{u=\frac{\ee{1}{2}{}\ee{3}{4}{}}{\ee{1}{3}{}\ee{2}{4}{}}\quad\quad\text{and}\quad\quad v=\frac{\ee{1}{4}{}\ee{2}{3}{}}{\ee{1}{3}{}\ee{2}{4}{}},}[EqCR]
in embedding space, the quantity
\eqna{
G_{\Delta_m}(u,v)&=c\frac{\ee{1}{2}{\frac{1}{2}(\Delta_i+\Delta_j)}\ee{1}{3}{\frac{1}{2}(\Delta_k-\Delta_\ell)}\ee{1}{4}{\frac{1}{2}(\Delta_i-\Delta_j-\Delta_k+\Delta_\ell)}\ee{3}{4}{\frac{1}{2}(\Delta_k+\Delta_\ell)}}{\ee{2}{4}{\frac{1}{2}(\Delta_i-\Delta_j)}}\tilde{G}\\
&=c\frac{\ee{1}{2}{\frac{1}{2}\Delta_m}\ee{1}{3}{\frac{1}{2}(\Delta_k-\Delta_\ell)}\ee{1}{4}{\frac{1}{2}(\Delta_i-\Delta_j-\Delta_k+\Delta_\ell)}\ee{3}{4}{\frac{1}{2}\Delta_m}}{\ee{2}{4}{\frac{1}{2}(\Delta_i-\Delta_j)}}\\
&\phantom{=}\hspace{1cm}\times\D(\eta_1,\eta_2)^{-\frac{1}{2}(\Delta_i-\Delta_j+\Delta_m)}\D(\eta_3,\eta_4)^{-\frac{1}{2}(\Delta_k-\Delta_\ell+\Delta_m)}\frac{1}{\ee{2}{4}{\Delta_m}},
}[EqG]
obtained from \eqref{EqtG}, is the scalar conformal block up to a normalization factor $c$.  The scalar conformal block \eqref{EqG} is thus encoded in the function
\eqn{G_d^{(p,q;s)}=\D(\eta_1,\eta_2)^p\D(\eta_3,\eta_4)^q\ee{2}{4}{s},}[EqGd]
for $p=-\frac{1}{2}(\Delta_i-\Delta_j+\Delta_m)$, $q=-\frac{1}{2}(\Delta_k-\Delta_\ell+\Delta_m)$ and $s=-\Delta_m$.

Acting first with $\D(\eta_3,\eta_4)$ in \EqGd leads to the following result
\eqn{G_d^{(p,q;s)}=(-2)^q(-s)_q(-s+1-d/2)_q\D(\eta_1,\eta_2)^p\ee{2}{3}{q}\ee{2}{4}{s-q},}[EqGdp]
as can be seen from the three-point correlation function \eqref{EqDiff12e23}.  Acting then with $\D(\eta_1,\eta_2)$ is not as straightforward.  First, it is important to notice that $\D(\eta_1,\eta_2)$ commutes with $\ee{1}{2}{\alpha}$ and acts non-trivially only on $\ee{2}{3}{}$ and $\ee{2}{4}{}$.  Moreover, the form of the conformal block must be expressible in terms of the conformal ratios \EqCR.  Thus, it is convenient to re-express the differential operator $\D(\eta_1,\eta_2)$ as
\eqn{\D(\eta_1,\eta_2)=\frac{\ee{1}{3}{}\ee{1}{4}{}}{\ee{1}{2}{}\ee{3}{4}{}}\D(u,v),}
where by the chain rule one obtains
\eqn{\D(u,v)=(-2)\left\{u^3\partial_u^2-u^2(u-\bar{v})\partial_u\partial_{\bar{v}}+u^2(1-\bar{v})\partial_{\bar{v}}^2-\left(\tfrac{d}{2}-2\right)u^2\partial_u-u\left[u+\left(\tfrac{d}{2}-1\right)\bar{v}\right]\partial_{\bar{v}}\right\}.}[EqD]
Here $\bar{v}=1-v$ for future convenience.  Re-expressing \eqref{EqGdp} in terms of the conformal ratios gives
\eqn{G_d^{(p,q;s)}=(-2)^q(-s)_q(-s+1-d/2)_q\frac{\ee{1}{3}{p+q-s}\ee{1}{4}{p-q}}{\ee{1}{2}{p-s}\ee{3}{4}{p-s}}\D(u,v)^pu^{-s}v^q,}[EqGdpp]
which leads to
\eqn{G_{\Delta_m}(u,v)=c(-2)^q(-s)_q(-s+1-d/2)_qu^{s/2-p}\D(u,v)^pu^{-s}v^q,}[EqGp]
for the scalar conformal block.  By a change of variables, it is possible to verify that the generating function $\D(u,v)^pu^{-s}v^q$ is related to Koornwinder polynomials \cite{Koornwinder197448,Koornwinder197459,Koornwinder1974357,Koornwinder1974370}.  Indeed, when $p$ is a positive integer, \cite{doi:10.1137/0509028} showed that
\eqn{\D(u,v)^pu^{-s}v^q=2^p(-q)_p(q-s)_pu^{2p-s}v^{q-p}F_4(-p,s-p+d/2;q-p+1,s-p-q+1;v/u,1/u),}
where $F_4$ is the fourth Appell function,
\eqn{F_4(\alpha,\beta;\gamma,\delta;x,y)=\sum_{m,n\geq0}\frac{(\alpha)_{m+n}(\beta)_{m+n}}{(\gamma)_m(\delta)_nm!n!}x^my^n.}
This solution, however, does not have the appropriate limiting behavior at $u\to0$ and $\bar{v}\to0$ and is furthermore not symmetric under the interchange of $p$ and $q$, contrary to \eqref{EqG}.  The correct answer is
\eqna{
\D(u,v)^pu^{-s}v^q&=(-2)^p(-s)_p(-s+1-d/2)_pu^{p-s}v^{q-p}\\
&\phantom{=}\hspace{1cm}\left[\frac{\Gamma(p-q)\Gamma(-s)}{\Gamma(p-s)\Gamma(-q)}F_4(q-s,-p;-s+1-d/2,q-p+1;u,v)\right.\\
&\phantom{=}\hspace{1cm}+\frac{\Gamma(q-p)\Gamma(-s)}{\Gamma(q-s)\Gamma(-p)}v^{p-q}F_4(p-s,-q;-s+1-d/2,p-q+1;u,v)\left.\right],
}[EqRodrigues]
which implies
\eqna{
G_{\Delta_m}(u,v)&=c(-2)^{p+q}(-s)_p(-s)_q(-s+1-d/2)_p(-s+1-d/2)_qu^{-s/2}v^{q-p}\\
&\phantom{=}\hspace{1cm}\left[\frac{\Gamma(p-q)\Gamma(-s)}{\Gamma(p-s)\Gamma(-q)}F_4(q-s,-p;-s+1-d/2,q-p+1;u,v)\right.\\
&\phantom{=}\hspace{1cm}+\frac{\Gamma(q-p)\Gamma(-s)}{\Gamma(q-s)\Gamma(-p)}v^{p-q}F_4(p-s,-q;-s+1-d/2,p-q+1;u,v)\left.\right].
}[EqGpp]
This solution now has the appropriate limiting behavior and is symmetric under the interchange of $p$ and $q$.  It is interesting to note that both terms in \eqref{EqRodrigues} satisfy the recurrence relation originating from the generating function $\D(u,v)^pu^{-s}v^q$ for integer $p$. However, only the first one has the correct boundary condition at $p=0$.  The second term is homogeneous and thus vanishes at $p=0$.  Thus, the second term can be added with an appropriate constant prefactor to enable the $p\leftrightarrow q$ symmetry.  Moreover, \eqref{EqGpp} can be seen as the non-compact version of the Koornwinder polynomials. Consequently, the orthogonality properties of the polynomials, when translated to the non-compact case, should allow an analytic version of the conformal bootstrap.

Finally, defining
\eqn{G(\alpha,\beta,\gamma,\delta;x,y)=\sum_{m,n\geq0}\frac{(\delta-\alpha)_m(\delta-\beta)_m}{(\gamma)_mm!}\frac{(\alpha)_{m+n}(\beta)_{m+n}}{(\delta)_{2m+n}n!}x^my^n,}[EqnGDO]
which can be related to the fourth Appell function with the help of a classical identity for the ${}_2F_1$ Gauss hypergeometric function,
\eqna{
G(\alpha,\beta,\gamma,\delta;x,y)&=\sum_{m\geq0}\frac{(\delta-\alpha)_m(\delta-\beta)_m}{(\gamma)_mm!}\frac{(\alpha)_m(\beta)_m}{(\delta)_{2m}}{}_2F_1(\alpha+m,\beta+m;\delta+2m;y)x^m\\
&=\frac{\Gamma(\delta)\Gamma(\delta-\alpha-\beta)}{\Gamma(\delta-\alpha)\Gamma(\delta-\beta)}F_4(\alpha,\beta;\gamma,\alpha+\beta-\delta+1;x,1-y)\\
&\phantom{=}\hspace{20pt}+\frac{\Gamma(\delta)\Gamma(\alpha+\beta-\delta)}{\Gamma(\alpha)\Gamma(\beta)}(1-y)^{\delta-\alpha-\beta}F_4(\delta-\alpha,\delta-\beta;\gamma,\delta-\alpha-\beta+1;x,1-y),
}
the scalar conformal block becomes
\eqn{G_{\Delta_m}(u,v)=u^{-s/2}v^{q-p}G(q-s,-p,-s+1-d/2,-s;u,\bar{v}).}
Here the constant $c$ was chosen to reproduce the conformal blocks of \cite{Dolan:2003hv}.

This subsection demonstrates that the OPE in embedding space allows the computation of the scalar conformal block.  Generalizing the OPE in embedding space to arbitrary fields leads to all arbitrary conformal blocks, which are computed as appropriate derivatives of the minimal scalar conformal block.\footnote{Recent work on arbitrary conformal blocks in 3 and 4 dimensions can be found in \cite{Rejon-Barrera:2015bpa,Iliesiu:2015akf,Echeverri:2016dun}, where the seed conformal blocks were obtained.  It is important to note that the OPE in embedding space implies that all conformal blocks are generated from the scalar conformal block.}  The details of the general case will be discussed elsewhere.

%%%%%%%%%%%%%%%%%%%%%%%%%%%%%%%%%%%%%%%%%%%%%%%%%%

\subsection{Conformal Bootstrap in Embedding Space}\label{SSecCB}

Repeating the previous section for the $t$- and $u$-channels, it is possible to implement the conformal bootstrap directly in embedding space.  Indeed, applying \eqref{EqFourPT} in all channels leads to the following consistency conditions,
\eqna{
\sum_{m,n}\sum_{a=1}^{N_{ijm}}\sum_{b=1}^{N_{k\ell n}}&\cOPE{a}{i}{j}{m}\cOPE{}{m}{n}{\1}\cOPE{b}{k}{\ell}{n}\DOPE{a}{i}{j}{m}(\eta_1,\eta_2)\DOPE{b}{k}{\ell}{n}(\eta_3,\eta_4)\DOPE{}{m}{n}{\1}(\eta_2,\eta_4)\\
&=(-1)^{F_{jk}}\sum_{m,n}\sum_{a=1}^{N_{ikm}}\sum_{b=1}^{N_{j\ell n}}\cOPE{a}{i}{k}{m}\cOPE{}{m}{n}{\1}\cOPE{b}{j}{\ell}{n}\DOPE{a}{i}{k}{m}(\eta_1,\eta_3)\DOPE{b}{j}{\ell}{n}(\eta_2,\eta_4)\DOPE{}{m}{n}{\1}(\eta_3,\eta_4)\\
&=(-1)^{F_{ik}}(-1)^{F_{i\ell}}\sum_{m,n}\sum_{a=1}^{N_{i\ell m}}\sum_{b=1}^{N_{jkn}}\cOPE{a}{i}{\ell}{m}\cOPE{}{m}{n}{\1}\cOPE{b}{j}{k}{n}\DOPE{a}{i}{\ell}{m}(\eta_1,\eta_4)\DOPE{b}{j}{k}{n}(\eta_2,\eta_3)\DOPE{}{m}{n}{\1}(\eta_4,\eta_3),
}[EqCB]
where $(-1)^{F_{ij}}$ takes into account the statistics of the permuted fields in the four-point correlation function.

Projecting onto different irreducible representations of the Lorentz group, eliminating the common tensor structures, and applying orthogonality conditions on the resulting equations lead to the implementation of the conformal bootstrap directly in embedding space.

%%%%%%%%%%%%%%%%%%%%%%%%%%%%%%%%%%%%%%%%%%%%%%%%%%
%%%%%%%%%%%%%%%%%%%%%%%%%%%%%%%%%%%%%%%%%%%%%%%%%%

\section{Discussion and Conclusion}\label{SecConc}

In this paper we showed how to calculate conformal blocks from the operator product expansion in embedding space.  We focused on the scalar exchange in a four-point correlation function of four scalar fields and reproduced the results found in the literature.  The power of our technique comes from the simplifications one encounters in embedding space where the conformal group acts linearly. In our approach,  all conformal blocks are computed from appropriate derivatives of the minimal scalar conformal block, and then the conformal bootstrap is executed in embedding space.

Indeed, since the embedding space is a more natural setting for a conformal field theory, there is absolutely no reason to project back onto position space to implement the conformal bootstrap.  It is more convenient to single out the contributions with the same structure under the Lorentz group in embedding space and thereafter proceed with the conformal bootstrap.

Since the building block for implementing the conformal bootstrap is the scalar conformal block, it is important to study its properties carefully.  Being related to Koornwinder polynomials, which are orthogonal polynomials in two variables, it is normal to expect that there exists a non-compact extension of the related orthogonality properties for the conformal blocks.  The crucial step from the trigonometric to the hyperbolic case has been recently made in \cite{Isachenkov:2016gim}, which relates the conformal blocks for symmetric-traceless field exchange in four-point correlation functions of four scalar fields to virtual Koornwinder polynomials (see \cite{ref1}).  This important observation allows an analytic version of the conformal bootstrap for four-point correlation functions of four scalar fields.  Since all conformal blocks are obtained from derivatives of the minimal scalar conformal block, it strongly suggests that the analytic conformal bootstrap can be extended completely to include fields in arbitrary representations of the Lorentz group.  Such a complete analytic conformal bootstrap would give a Jacobi-like definition of conformal field theories where only the conformal data (the conformal dimensions, the irreducible representations of the Lorentz group and the operator product expansion coefficients) appear, without any conformal blocks.

A supersymmetric extension of the operator product expansion in embedding space to the operator product expansion in supersymmetric embedding space (see \cite{Goldberger:2011yp,Goldberger:2012xb}) is also an interesting future topic of research, especially when combined with a possible supersymmetric extension of the Casimir equations and their solutions as hinted in \cite{Isachenkov:2016gim}.  This would also allow a complete extension of the analytic superconformal bootstrap to all superfields, resulting again in a definition of superconformal field theories in terms of the superconformal data only.

In a forthcoming publication it will be shown how the different differential operators $\DOPE{a}{i}{j}{k}(\eta_1,\eta_2)$ for arbitrary fields in the OPE in embedding space are built from a unique differential operator.

%%%%%%%%%%%%%%%%%%%%%%%%%%%%%%%%%%%%%%%%%%%%%%%%%%
%%%%%%%%%%%%%%%%%%%%%%%%%%%%%%%%%%%%%%%%%%%%%%%%%%

\ack{
The authors would like to thank A.~Liam Fitzpatrick, Vincent Genest, Walter D.~Goldberger, Tom H.~Koornwinder, Pierre Mathieu, Gregory W.~Moore, and Vyacheslav S.~Rychkov for useful discussions.  The authors are indebted to CERN where the original idea behind this work was born.  The work of JFF is supported by NSERC.  WS is supported in part by the U.~S.~Department of Energy under the contract DE-FG02-92ER-40704.
}

%%%%%%%%%%%%%%%%%%%%%%%%%%%%%%%%%%%%%%%%%%%%%%%%%%
%%%%%%%%%%%%%%%%%%%%%%%%%%%%%%%%%%%%%%%%%%%%%%%%%%

\bibliography{Short-Final}

\end{document}